\documentclass[12pt]{iopart}
\usepackage{graphicx}
\usepackage{amssymb}

\def\prd{Phys. Rev. D}

\def\w{\omega}

\def\a{\alpha}

\def\th{t_0}
\def\*{\cdot}

\def\d{\partial}
\def\de{\delta}

\def\ba#1{\overline{#1}}
\def\v#1{\mathbf{#1}}
\def\uv#1{\underline{\mathbf{#1}}}

\def\ra{\rightarrow}
\def\ax{\uv{x}}
\def\ay{\uv{y}}
\def\PD{{\bf PD}}

\def\X{\ax}
\def\s{ \sigma}
\def\Ss{\mathcal S}

\begin{document}

\title{Balanced homodyne detectors and Casimir energy densities}

\author{P Marecki}

\address{ITP, Universitaet Leipzig, Postfach 100 920, D-04009 Leipzig, Germany}
\ead{pmarecki@gmail.com}

\begin{abstract}
We recall and generalize the analysis of the output of the so-called balanced homodyne detectors. The most important feature of these detectors is their ability to quantify the vacuum fluctuations of the electric field, that is expectation values of products of (quantum-) electric-field operators. More precisely, the output of BHDs provides information on the one- and two-point functions of arbitrary states of quantum fields. We compute the two-point function and the associated spectral density for the ground state of the quantum electric field in a Casimir geometry. Furthermore, we predict a  position- and frequency-dependent pattern of BHD responses if a device of this type were to be placed between Casimir plates. This points to a potential new characterization of ground states in Casimir geometries, which would not only complement the current global methods (Casimir forces), but also improve understanding of sub-vacuum energy densities present in some regions in these geometries.

\end{abstract}

\maketitle

\section{Introduction}

One of the most intriguing predictions of quantum field theory is that expectation values of classically positive quantities are not necessarily positive. The far-reaching impli\-cations of this are clear in the case of energy-\emph{densities} of quantum fields. These densities are classically positive but there might be regions in spacetime, where the expectation values of the corresponding quantum operators are negative (sub-vacuum). While a number of surprising results related to this phenomenon have been discovered\footnote{See, for example, \cite{FR,FGO,appl,DO} and the references therein.}, attempts were also made to see negative energy densities (related to vacuum fluctuations) in experiments. Devices capable of detecting and quantifying the vacuum fluctuations were first proposed in the context of quantum optics by Chen and Yuan \cite{first}, and are known as balanced homodyne detectors\footnote{See \cite{pulsed} for a modern experimental realization of a BHD.}. With their help it was confirmed that squeezed states of light  exhibit regions of sub-vacuum electric field fluctuations (see e.g. \cite{squeezed}).

Sub-vacuum fluctuations, i.e. expectation values of the square of the electric field, appear also for ground states of quantum fields under influence of static external conditions. In this paper we ask the question, whether these interesting properties of ground states can be detected experimentally. We analyze the response of a hypothetical BHD-type device placed between Casimir plates, and derive a prediction for the expectation value of the variance of its current. We show, that this variance is related in a simple way to the spectral function of the Casimir ground state, and therefore argue that a BHD-measurement would provide a detailed characterization of this state,  complementary to the currently available experimental results restricted so far to  \emph{global} quantities such as Casimir \emph{forces} \cite{parallel}.

\section{Balanced homodyne detector under influence of (static) external conditions}
In this section we recall the general description of photodiodes and balanced homodyne detectors appropriate in the case of quantum fields under influence of static external conditions. Subsequently we exhibit a relation between the variance of BHD-output and the so-called spectral densities which characterize two-point functions of ground states.

\subsection{Photodiodes and BHDs}
We model photodetection as a process in which a simple quantum system, e.g. an electron, initially in a well-localized bound state gets excited to the continuum of scattering states. The excitation is caused by the (dipole) interaction with the quantum electric field. Employing the first-order time-dependent perturbation theory one arrives at the following expression for the probability of excitation:
\begin{equation}
   \PD(g,\ax)=\int_{-\infty}^{\infty} d\tau ds\, g(\tau)g(s)\,
    G^{ij}(\tau-s)\, \left\langle E_i(\tau,\ax)\, E_j(s,\ax)\right\rangle_\Ss,
\end{equation}
where $E_i(\tau, \ax)$ is the operator of the quantum electric field \emph{at the point} $\ax$ where the electron was initially essentially localized, $\langle \ldots \rangle_\Ss$ stands for the expectation value w.r.t. the initial state of the quantum field, $G^{ij}(\tau - s)$ is the electronic correlation function (depending only on its initial and final states\footnote{See complement $A_{II}$ of \cite{CCT} or \cite{PM} for concrete expressions.}), and $g(\tau)$ is a test function supposed to be equal to one during the measurement and vanishing (smoothly) elsewhere.

A balanced homodyne detector consists of an arrangement of two photodiodes, whose outputs are subtracted, illuminated with an auxiliary coherent state, $F$, of the radiation field (see Fig. 1). The expectation value of the observable corresponding to the charge collected at $P$ is the difference of excitation probabilities of the photodiodes:
\begin{equation}
\langle J\rangle_\Ss=\PD(g,\ax)-\PD(g,\ay).
\end{equation}

\begin{figure}[ht]\centering
  \includegraphics[scale=0.7]{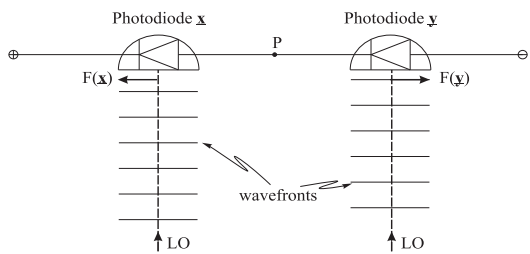}
  \caption{BHD with a LO.  Note, that the setup is arranged in such a way, that the electric field of the LO at $\ax$ has a reversed direction w.r.t. that at $\ay$.}
  \label{bhd}
\end{figure}

The coherent state $F$, called the ``local oscillator'', is used only as a tool to investigate properties of a certain state $\Ss$ of the quantum radiation field. On a BHD the state $\Ss$ is  ``blended'' with $F$, which we describe as follows:
\begin{equation}
  \langle P[E_i(t,\v x)] \rangle_{(\Ss,F)}=\langle P[F_i(t,\v x)+E_i(t,\v
  x)]\rangle_\Ss,
\end{equation}
where $F_i(t,\v x) = \langle E_i(t,\v x)\rangle_F$ denotes the electric field of the LO, and $(\Ss,F)$ is the state resulting from the ``blending''. The symbol $P$ stands for an arbitrary QFT-observable (Polynomial) constructed out of the  electric field operators.

Usually balanced homodyne detectors are described in terms of the quantum electric field operators in the vacuum representation. Here we shall employ the more general description \cite{PM}, which is applicable to the case of quantum fields under the influence of static external conditions. The electric field operator possesses the decomposition
\begin{equation}
    E_i(t,\ax)=\int d\nu(p^a) \left [e^{-i\w_p t}\, \psi_i(p^a,\ax) b(p^a)+e^{i\w_p t}\, \ba{ \psi_i(p^a,\ax)} b^*(p^a)\right].
\end{equation}
where the multi-index $p^a$ contains the parameters of electromagnetic waves supported by the environment/medium (such as types of waves and their wave vectors), $\psi_i(p^a, \ax)$ denotes the electric field of the solution of the Maxwell equations with these parameters and $d\nu(p^a)$ denotes the appropriate measure\footnote{In vacuum, $p^a=(\vec p,\a)$ where $\vec p\in  \mathbb R^3$ is the wave vector, $\a=1,2$ denotes the transversal polarizations and $e^{-ic|\vec p| t}\psi_i(p^a,\ax)$ corresponds to the appropriately normalized plane waves.}. The frequency, $\w(p^a)$, is given by a known dispersion relation. The functions $\psi_i(p^a, \ax)$ are normalized in a way compatible with the relation  $[b(p^a),b^*(k^a)]=\de(p^a-k^a)$.

The LO is an almost monochromatic wave packet which at the point $\ax$ can be expanded into
\begin{equation}
F_i(\tau,\ax )=\int d\w k_i(\w) \left[e^{-i\w ( \tau-\th)}+e^{i\w( \tau-\th)} \right],
\end{equation}
where the real vector-valued function of frequencies, $k_i(\w)$,  is sharply concentrated around $\w=\w_{LO}$, called the frequency of the local oscillator. The parameter $\th$ allows for a shift of the phase of the LO field.

The detector is arranged (balanced) is such a way that $F_i(\tau,\ax )=-F_i(\tau,\ay )$ (see Fig. 1). Under this condition it can be shown \cite{PM} that in the limit $g\ra 1$ (even under the influence  of external conditions) the leading contribution to the expectation value of the BHD current is given by
\begin{equation}\label{Jexpect}
    \langle J\rangle_{(\Ss,F)}=A(\w_{LO})\Big\langle  E(\th,\ax)|_{k(\w)}+ E(\th,\ay)|_{k(\w)}\Big\rangle_\Ss,
\end{equation}
where
\begin{equation}\label{Ek}
\hspace*{-1.5cm}E(\th,\ax)|_{k(\w)}=\int d\nu(p^a)k^i(\w_p)\left [e^{-i\w_p \th}\, \psi_i(p^a,\ax) b(p^a)+e^{i\w_p \th}\, \ba{ \psi_i(p^a,\ax)} b^*(p^a)\right],
\end{equation}
and the constant $A(\w_{LO})$ is the only trace of the electronic correlation function $G^{ij}(\tau - s)$. Note, that the balancing of the detector canceled the term scaling quadratically with the amplitude of the LO field.

If the expectation value of the electric field operator (and thus the leading term of $\langle J\rangle_{(\Ss,F)}$) vanishes, then the variance of the BHD-output is $\langle J^2\rangle_{(\Ss,F)}$ and, as we shall see, provides a characterization of the two-point function of the state $\Ss$. For the leading term of this variance we find \cite{PM}
\begin{equation}\label{J_squared}
\hspace*{-2.0cm}\langle J^2 \rangle_\Ss=A^2(\w_{LO}) \Big\langle  \left[E(\th,\ax)|_{k(\w)}+E(\th,\ay)|_{k(\w)}\right] \left[ E(\th,\ax)|_{k(\w)}+E(\th,\ay)|_{k(\w)}\right]\Big\rangle_\Ss.
\end{equation}
This quantity scales quadratically with the amplitude and thus linearly with the power of the local oscillator field; by performing measurements with different powers of the LO  the two-point functions can be quantitatively estimated.

\subsection{Relation between the BHD output and spectral densities}
Let us now specify $\Ss$ to be the ground state (denoted by $G$ in the sequel) of the quantum field under the influence of static external conditions. In this case the one-point function vanishes and the two-point function possesses the representation
\begin{equation}
 \langle E_i(t,\ax) E_i(\tau,\ay) \rangle_G=\int d\nu(p^a)\, e^{-i\w_p(t-\tau)}\, \psi_i(p^a,\ax) \ba{\psi_j(p^a,\ay)}.
\end{equation}
Let us introduce the spectral density which is defined as the Fourier transform of the two-point function $\langle E_i(t,\ax) E_i(0,\ay) \rangle_G$ w.r.t. time $t$. We find
\begin{equation}
 \sigma_{ij}(\w,\ax,\ay)=\int d\nu(p^a) \delta(\w_p-\w)\, \psi_i(p^a,\ax) \ba{\psi_j(p^a,\ay)}.
\end{equation}
In the expression for the variance of the output of the BHD for $G$, Eq. (\ref{J_squared}), there appear (time-independent) factors
\begin{equation}
 R(\ax,\ay)\equiv\big \langle E(\th,\ax)|_{k(\w)}\, E(\th,\ay)|_{k(\w)}\big \rangle_G,
\end{equation}
which simplify to
\begin{equation}
  R(\ax,\ay)=\int d\nu(p^a)\, k^i(\w_p)k^j(\w_p)\,  \psi_i(p^a,\ax)\ba{\psi_j(p^a,\ay)}.
 \end{equation}
It can be easily seen that the the function $R(\ax,\ay)$ is just the spectral density smeared in frequency by the product of amplitudes of the local oscillator, $k^i(\w)k^j(\w)$:
\begin{equation}\label{correspondence}
 R(\ax,\ay)=\int d\w\, k^i(\w)k^j(\w)\, \sigma_{ij}(\w,\ax,\ay).
\end{equation}
In this way, by exploring the freedom of choosing the locations of the photodiodes as well as polarizations, phases and frequencies of the LO one can obtain a detailed characterization of one- and two-point functions of any state $\Ss$ of the quantum electric field. This is of fundamental importance for QFT under influence of external conditions.

\section{Spectral densities of ground states in Casimir geometries}
\label{Casimir}
Here we shall derive the spectral density  for the quantum electric field between two parallel, perfectly conducting plates (at $x=0$ and $x=a$). The diagonal part of this density, $\sigma_{yy}(\w,\ax,\ay)$ can be found in the paper of Hacyan et al.,  \cite{HJSV}.

The ground-state two-point function, $\langle E_i(x^a)E_j(\tilde x^a)\rangle_G$, can be computed (cf. Eq. (2.25c) of \cite{HJSV}) by  differentiating the image-sums
\begin{equation}
\hspace*{-1cm}F^{\mp}(x,\tilde x)=-\frac{1}{4\pi^2}\sum_{n=-\infty}^\infty \frac{1}{(x\mp \tilde x-n L)^2+(y-\tilde y)^2+(z-\tilde z)^2-(t-\tilde t)^2}
\end{equation}
where $x^a=(t,x,y,z)$ and $L=2a$ is \emph{twice} the distance between the plates. Specifically, for the $y$-components of the electric field operators we have
\begin{equation}
\langle E_y(x^a)E_y(\tilde x^a)\rangle_G=(\d_x^2+\d_z^2)\left[F^-(x^a,\tilde x^a)-F^+(x^a,\tilde x^a)\right],
\end{equation}
where differentiations are performed w.r.t. $x^a$ only. Here, we will be interested in the spectral density for $x^a=(s,\ax)$, $\tilde x^a=(0,\ay)$ with $\ax=(x,0,0)$ and $\ay=(x,y,0)$, where $x\in[0,a]$ and  $y\in\mathbb R$.  We find
\begin{equation}\label{2pt}
 \hspace{-2.5cm}\langle E_y(\ax) E_y(\ay)\rangle_G=\frac{1}{\pi^2}\sum_{n=-\infty}^{n=\infty}\left[\frac{A^2+s^2}{(s^2-A^2)^3}-
\frac{B^2+s^2}{(s^2-B^2)^3}\right]+\frac{2y^2}{\pi^2}\left[
\frac{1}{(s^2-B^2)^3}-\frac{1}{(s^2-A^2)^3}
\right],
\end{equation}
where
\begin{equation}
 A=\sqrt{(nL)^2+y^2}, \qquad
 B=\sqrt{(2x-nL)^2+y^2}.
\end{equation}
We now compute the spectral density by performing the Fourier transform and get
\begin{equation}
 \hspace{-1.5cm}\s_{yy}(\w,\ax,\ay)=\frac{\w^3}{4\pi^2}\sum_n\left\{
\Bigl[Q(\w A)-Q(\w B)\Bigr]+y^2\left[\frac{1}{B^2}\, W(\w B)
-\frac{1}{A^2}\, W(\w A)\right]
\right\},
\end{equation}
with
\begin{equation}
 Q(x)=\frac{\sin x }{x}+\frac{\cos x }{x^2}-\frac{\sin x}{x^3},\qquad
 W(x)=\frac{\sin x }{x}+\frac{3 \cos x }{x^2}-\frac{3\sin x}{x^3}.
\end{equation}
This simplifies to
\begin{equation}\label{sG}
 \s_{yy}(\w,\ax,\ax)=\frac{\w^3}{4\pi^2}\sum_n\left[
Q(\w nL )-Q(\w |2x-nL|)\right]
\end{equation}
for $y=0$, and coincides with Eq. (2.30) of \cite{HJSV}. Note, that the $n=0$ term of the A-sum in Eq. (\ref{2pt}) corresponds to the vacuum spectral density (i.e. to the situation without any external conditions),
\begin{equation}
  \s^\Omega_{yy}(\w,\ax,\ay)=\frac{\w^3}{2\pi^2}\left[\frac{\sin(\w y)}{(\w y)^3}-\frac{\cos (\w y)}{(\w y)^2}\right],
\end{equation}
which is regular at $y=0$, with $\s^\Omega_{yy}(\w,\ax,\ax)=\w^3/6\pi^2$. For large distances between $\ax$ and $\ay$ the vacuum spectral density falls as $1/y^2$.
\begin{figure}[h]\centering
  \includegraphics[scale=0.45]{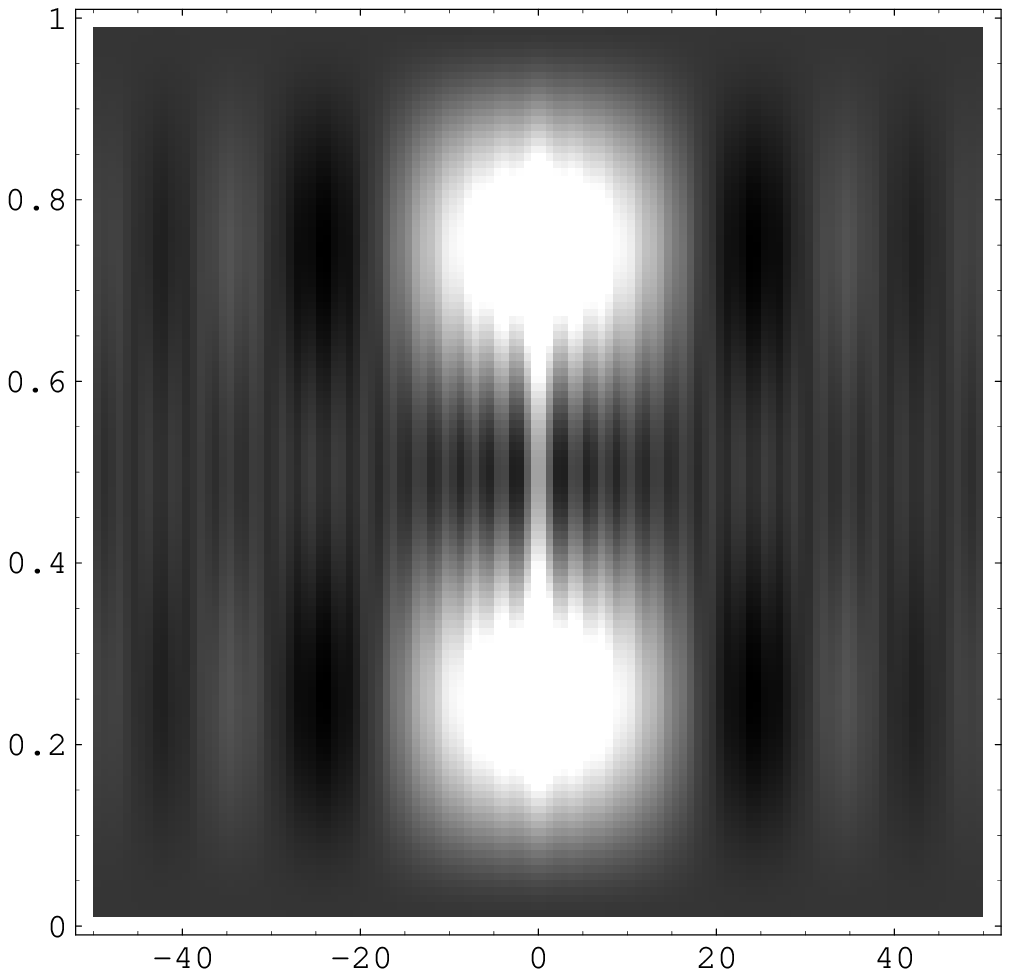}   \includegraphics[scale=0.7]{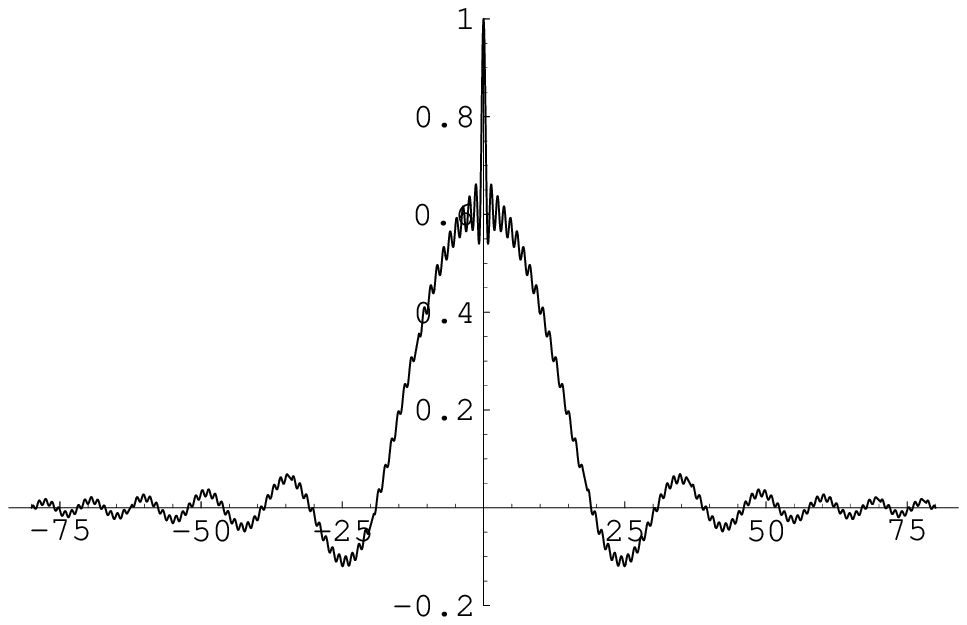}
\caption{ [Left:] The spectral density  $\s_G(\w,\ax,\ay)$ for $\ax=(x,0,0)$, $\ay=(x,y,0)$ as a function of $x\in[0,a]$ and $y\in[-50a,50a]$.   [Right:]
Normalized spectral density $\s_G(\w,\ax,\ay)/\s_G(\w,\ax,\ax)$ for $y\in[-50a,50a]$ at $x=0.75a$ (symmetrically including $1000$ terms of the $n$-sums). All distances are in $\mu$m, the frequency is $\w=2\pi c/a$ and the plate separation $a=1\mu$m.}
\end{figure}
We present numerical evidence (see Fig. 2) for a similar behavior in the case of ground state in Casimir geometry. The diagonal part of spectral density $\s_{yy}(\w,\ax,\ax)$, which can be computed from (\ref{sG}), and the possibility of its experimental detection will be discussed in the next section.

\section{Blueprint for the proposed experiment}
Let us discuss the possibility of an experimental detection of the  Casimir spectral density with a BHD-type device. For plates separated by $a=1\mu$m,  photodiodes of submicrometer width ($x$-direction) and submilimeter length ($y$-direction) would need to be used\footnote{Photodiodes  of this size have already been constructed, and their high-quantum-efficiency versions are under development, see \cite{sub_det} and references therein.}.
\begin{figure}[h]\centering
  \includegraphics[scale=0.6]{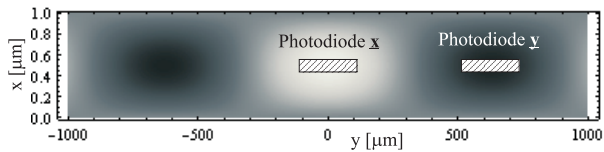}
\caption{Experimental setup drawn on the plot of the $F_y(\ax)$, the y-component of the electric field of  TE$_1$ mode of the Casimir cavity. The mode, serving as the LO, propagates in the $z$-direction perpendicular to the plot.}
\end{figure}
As shown in \mbox{Fig. 3} a coherent state in the TE$_1$ mode of the Casimir cavity with a \emph{very small wave number $p$} in $y$ direction would provide an appropriate local-oscillator. The Hertz potential of this mode is
\begin{equation}
 m(x^a)=\sin(k z-\w t)\cos({\pi x}/{a})\cos(p y)
\end{equation}
which gives the electric field $F_y$ much larger (in amplitude) than $F_x$:
\begin{eqnarray}
 F_x&=&-\frac{1}{c}\,\d_t\d_y m =\frac{-\w p}{c}\cos(k z-\w t)\cos({\pi x}/{a})\sin(p y)\\
 F_y&=&\quad\ \frac{1}{c}\d_t \d_x m =\ \frac{\w \pi}{ca}\cos(k z-\w t)\sin({\pi x}/{a})\cos(p y)
\end{eqnarray}
and the dispersion relation $\frac{\w^2}{c^2}=\frac{\pi^2}{a^2}+p^2+k^2$.
A BHD with such a LO and the photodiodes located as in Fig. 3 would by Eqs. (\ref{Jexpect},\ref{Ek}) be sensitive only to the $y$-component of the quantum electric field.  The variance of the detector's output, Eq. (\ref{J_squared}), contains four terms each of which is related to the spectral density by Eq. (\ref{correspondence}). As argued in the previous section the diagonal terms will be dominant if the photodiodes are separated by a sufficiently large distance\footnote{For instance for $y\gtrsim 40 a$ at $\w=2\pi c/a$ we observe $|\s_G(\w,\ax,\ay)/\s_G(\w,\ax,\ax)|<10\%$.}.
We expect the variance of the BHD-output to be
\begin{equation}
\langle J^2 \rangle_\Ss\approx2 A^2(\w_{LO})\Big\langle  E(\th,\ax)|_{k(\w)}E(\th,\ax)|_{k(\w)}\Big\rangle_G,
\end{equation}
with the expectation value on the RHS essentially equal to $\s_{yy}(\w_{LO},\ax,\ax)$.
\begin{figure}[h]\centering
  \includegraphics[scale=0.45]{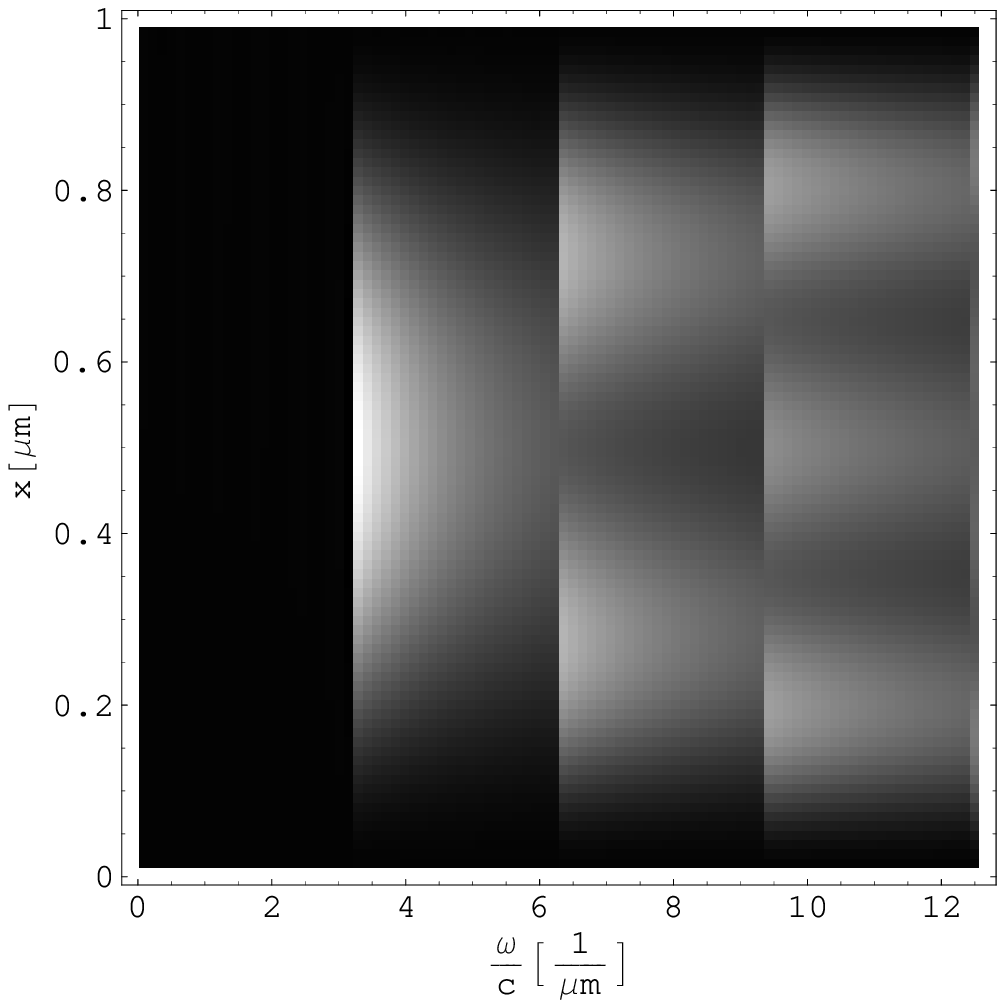}\quad   \includegraphics[scale=0.65]{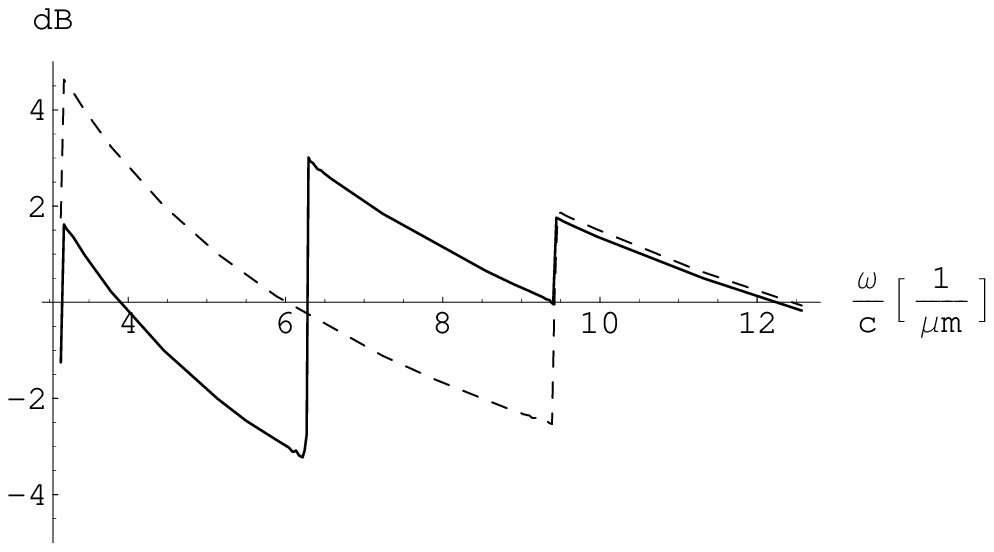}
\caption{ [Left:] Casimir spectral density normalized by vacuum spectral density,  $[{\s_{yy}(\w,\ax,\ax)}-{\s^\Omega_{yy}(\w,\ax,\ax)}]/\s^\Omega_{yy}(\w,\ax,\ax)$ as a function of the distance $x\in[0,a]$. Negative (sub-vacuum) values are black.  [Right:]
Suppression of vacuum fluctuations (see text) in dB for $x=0.25a$ (solid) and $x=0.5a$ (dashed).
Frequency range is $\w\in[0,4\pi c/a]$; $a=1\mu$m is assumed.}
\end{figure}
\mbox{Fig. 4} contains a plot of this spectral density\footnote{We note, that for $\w <\pi c/a$ the Casimir spectral density vanishes: $\s(\w,\X,\X)=0$, while discontinuities in it appear at $\w =n\pi c/a$.} as a function of the distance from the plates $x$ and the frequency $\w$.
For comparison with quantum-optical literature we have plotted the normalized difference between vacuum and ground-state spectral density,  $[{\s_{yy}(\w,\ax,\ax)}-{\s^\Omega_{yy}(\w,\ax,\ax)}]/\s^\Omega_{yy}(\w,\ax,\ax)$, and the ``suppression of vacuum fluctuations'',  $10\cdot\log_{10}\left[{\s(\w,\X,\X)}/{\s^\Omega(\w,\X,\X)}\right]$.

\section{Discussion}

A combination of arguments presented in this paper points to the possibility of experimental detection of spectral densities associated with quantum fields under influence of static external conditions. These densities contain the essence of the surprising phenomena associated with quantum fields in this regime, for instance by being related to negative (sub-vacuum) energy densities.

By recalling the analysis of the response of balanced homodyne detectors \cite{PM} general enough for our purposes, and computing the two-point functions and spectral densities for  the quantum electric field in a Casimir geometry we have estimated what a BHD-type device would measure if placed between Casimir plates. Position- and frequency-dependent pattern of BHD responses is predicted (Fig. 4). This pattern is static i.e. independent of the LO phase  and in some regions  corresponds to the suppression of vacuum fluctuations by at least $3$dB. Such a behavior is forbidden (by theorems known as quantum energy inequalities \cite{FR,appl}) for quantum fields without external conditions.
We therefore point to a possible test of yet unexplored generic quantum field theoretical effects in Casimir geometries, complementary to measurements of Casimir forces, and hope that experimental attempts to verify predictions derived here will follow.

\ack
I would like to thank K. Fredenhagen for encouragement and  T. Roman and L. Ford for discussions and for drawing my attention to the references \cite{HJSV,F}.


\section*{References}
\begin{thebibliography}{10}

\bibitem{FR}
 {Fewster} C J and  {Roman} T A 2003
\newblock {\em \prd}, 67:044003

\bibitem{FGO}
Ford L H , Grove P G  and  Ottewill A C 1992 \newblock {\em \prd}, 46:4566,


\bibitem{appl}
Marecki P  2002 \newblock {\em Phys. Rev. A}, 66:053801

\bibitem{DO}
Davies P C  and Ottewill A C  2002  \newblock {\em \prd}, 65:104014


\bibitem{first}
 {Yuen} H P and  {Chan} V W S 1983
\newblock {\em Optics Letters}, 8:177

\bibitem{pulsed}
{Hansen} H et al.  2001
\newblock {\em Optics Letters}, 26:1714

\bibitem{squeezed}
{Schneider} K et al.  1998
\newblock {\em Optics Express}, 3:59

\bibitem{parallel}
{Bressi} G et al.  2002
\newblock {\em Phys. Rev. Lett.},  88:041804


\bibitem{CCT}
{Cohen-Tannoudji} C, {Dupont-Roc} J and {Grynberg} G 1998
\newblock {\em {Atom-Photon Interactions: Basic Processes and Applications}}.
\newblock Wiley-VCH


\bibitem{PM}
Marecki P  2008 \newblock {\em Phys. Rev. A}, 77:012101


\bibitem{HJSV}
Hacyan S et al. 1990 \newblock {\em J. Phys. A}, 23:2401

\bibitem{sub_det}
Collin, S et al. 2003 \newblock {\em Appl. Phys. Lett.}, 83:1521

\bibitem{F}
Ford L H  1988 \newblock {\em Phys. Rev. D}, 38:528





\end{thebibliography}
\end{document}